\documentclass[aip,apl,reprint]{revtex4-1}
\usepackage{amsmath}
\usepackage{graphicx}
\usepackage{textcomp}
\usepackage{tikz}
\usepackage[hidelinks]{hyperref}
\usepackage{url}

\begin{document}

\title{Dual mode microwave deflection cavities for ultrafast electron microscopy}

\author{J.F.M.~van Rens}
\email{j.f.m.v.rens@tue.nl}
\affiliation{Department of Applied Physics, Coherence and Quantum Technology Group, Eindhoven University of Technology, P.O.~Box 513, 5600 MB Eindhoven, the Netherlands}
\author{W.~Verhoeven}
\affiliation{Department of Applied Physics, Coherence and Quantum Technology Group, Eindhoven University of Technology, P.O.~Box 513, 5600 MB Eindhoven, the Netherlands}
\author{E.R.~Kieft}
\affiliation{Thermo Fisher Scientific, Achtseweg Noord 5, 5651 GG Eindhoven, the Netherlands}
\author{P.H.A.~Mutsaers}
\affiliation{Department of Applied Physics, Coherence and Quantum Technology Group, Eindhoven University of Technology, P.O.~Box 513, 5600 MB Eindhoven, the Netherlands}
\author{O.J.~Luiten}
\affiliation{Department of Applied Physics, Coherence and Quantum Technology Group, Eindhoven University of Technology, P.O.~Box 513, 5600 MB Eindhoven, the Netherlands}
\affiliation{Institute for Complex Molecular Systems, Eindhoven University of Technology, P.O.~Box 513, 5600 MB Eindhoven, the Netherlands}
\date{\today}

\begin{abstract}
This paper presents the experimental realization of an ultrafast electron microscope operating at a repetition rate of 75 MHz based on a single compact resonant microwave cavity operating in \emph{dual mode}. This elliptical cavity supports two orthogonal TM$_{110}$ modes with different resonance frequencies that are driven independently. The microwave signals used to drive the two cavity modes are generated from higher harmonics of the same Ti:Sapphire laser oscillator. Therefore the modes are accurately phase-locked, resulting in periodic transverse deflection of electrons described by a Lissajous pattern. By sending the periodically deflected beam through an aperture, ultrashort electron pulses are created at a repetition rate of 75 MHz. Electron pulses with $\tau=(750\pm10)$ fs pulse duration are created with only $(2.4\pm0.1)$ W of microwave input power; with normalized rms emittances of $\epsilon_{n,x}=(2.1\pm0.2)$ pm rad and $\epsilon_{n,y}=(1.3\pm0.2)$ pm rad for a peak current of $I_p=(0.4\pm0.1)$ nA. This corresponds to an rms normalized peak brightness of $B_{np,\textrm{rms}}=(7\pm1)\times10^6$ A/m$^2$ sr V, equal to previous measurements for the continuous beam. In addition, the FWHM energy spread of $\Delta U = (0.90\pm0.05)$ eV is also unaffected by the dual mode cavity. This allows for ultrafast pump-probe experiments at the same spatial resolution of the original TEM in which a 75 MHz Ti:Sapphire oscillator can be used for exciting the sample. Moreover, the dual mode cavity can be used as a streak camera or time-of-flight EELS detector with a dynamic range $>10^4$.
\end{abstract}


\maketitle 

\section{Introduction}\label{sec:introduction}

Since the introduction of laser-triggered photo-cathodes in electron microscopy the field of single shot\cite{lagrange2008nanosecond} and stroboscopic\cite{zewail2010four} time-resolved electron imaging, diffraction and spectroscopy has resulted in many new insights in physics, chemistry, biology and material science.\cite{Vanacore2016, flannigan2018atomic} In a recent exciting development in which the flat photo-cathode is replaced by a nano-tip, the peak brightness of pulsed electron beams has become similar to values reported for conventional continuous Schottky field emitters.\cite{ehberger2015highly,Feist2017} The possibility of performing high-quality electron imaging, diffraction and spectroscopy with nanosecond to femtosecond temporal resolution causes the ultrafast transmission electron microscope (UTEM) to be one of the most powerful research tools to study ultrafast dynamics in the nanoscale world.

\begin{figure}
    \centering
    \includegraphics[width=0.5\textwidth]{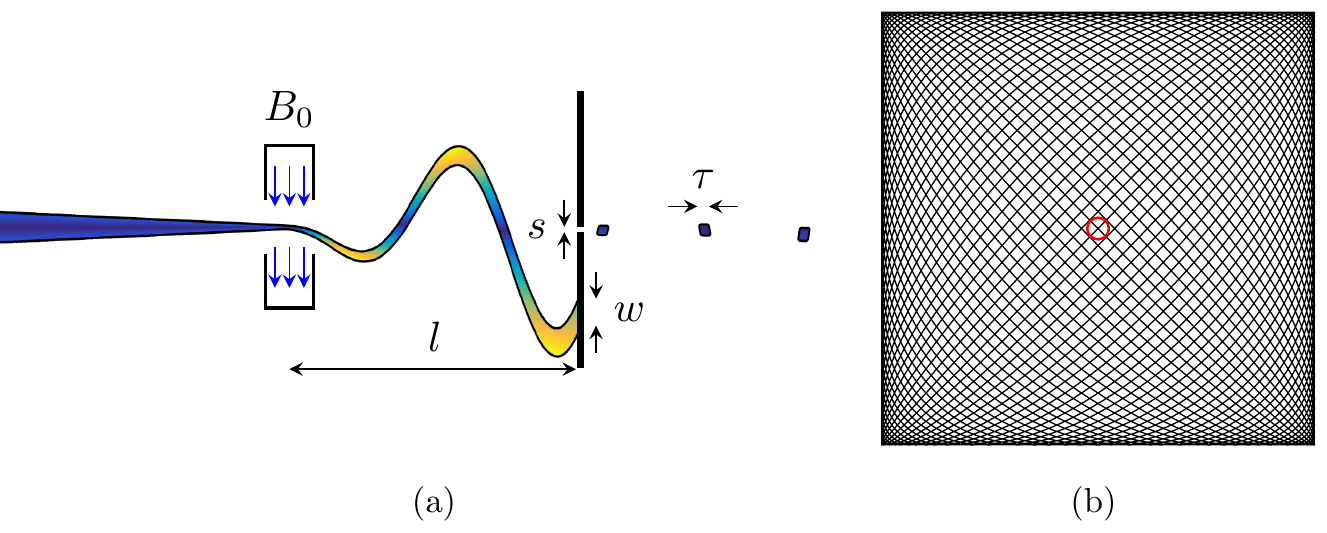}
    \caption{(a) Principle of beam chopping. The continuous beam is swept back and forth over an aperture, creating electron pulses at two times the resonance frequency. Because the two pulses create in each microwave period exit the aperture at slightly different angles, one beam can diverted back to the optical axis using standard deflection coils while the other beam is blocked by the Selected Area (SA)-aperture downstream. Parameters are defined in section \ref{sec:introduction}. \emph{Copyright Wiley-VCH Verlag GmbH \& Co. KGaA. Reproduced with permission.\cite{Verhoeven2018high}} (b) Parametric plot of two signals $x(t)=\sin{\omega_1 t}$ and $y(t)=\sin{(\omega_2 t+\delta\phi)}$ in which $\omega_1=2\pi \times 3$ GHz, $\omega_2=2\pi \times 3.075$ GHz and $\delta\phi=0$ for $t\in [0,13$ ns]. The resulting Lissajous figure describes the transverse shape of a beam of electrons deflected by a dual mode cavity in which the orthogonal modes are driven independently by two phase-locked signals of frequencies $f_1=3$ GHz and $f_2=3.075$ GHz. When a small aperture (red circle) is placed in the center of this Lissajous beam, two beams of ultrashort electron pulses are created, resulting in a combined repetition rate of $2\times75$ MHz. Because the two beams leave the chopping aperture at slightly different angles, one of them can be redirected to the optical axis using standard deflection coils, while the other one is blocked by the SA-aperture downstream. Alternatively the chopping aperture can be placed slightly off-center such that only one of the beams goes through, also reducing the repetition rate to 75 MHz.}
    \label{fig:Lissajous}
\end{figure}

Notwithstanding some truly impressive results,\cite{feist2015quantum,echternkamp2016ramsey} there are some drawbacks to photo-emission based sources. Bulky and expensive amplified laser systems are needed to create electrons, serious alterations are needed to the gun and inevitable laser pointing instabilities result in fluctuations in electron beam current. A promising alternative to photo-emission based pulsed beams involves chopping a continuous electron beam by the combination of a resonant microwave deflection cavity and a small aperture.\cite{Lassise2012compact, VanRens2018, Verhoeven2018high} Microwave cavities are specifically tailored metallic structures, in which oscillating electromagnetic fields with high amplitudes can be generated requiring only modest microwave input powers by driving the cavity at its resonance frequency. For example the TM$_{110}$ mode supports an oscillating magnetic field of amplitude $B_0$, oriented perpendicular to the symmetry axis of a cylindrical cavity. A beam of electrons passing through along this axis is transversely deflected at the resonance frequency of the cavity, as shown in figure \ref{fig:Lissajous}a. In combination with a small aperture of diameter $s$ positioned at a distance $l$ downstream, the beam is chopped to ultrashort pulses of duration
\begin{equation}\label{eq:tau}
\tau = \frac{\gamma m (s+w)}{4 q B_0 l},
\end{equation}
in which $\gamma$ the Lorentz factor, $m$ is the electron mass, $q$ the elementary charge and $w$ the diameter of the electron beam at the position of the chopping aperture.\cite{Verhoeven2018high} For 200 kV electrons and typical parameters $w=s=10$ $\mu$m and $l=10$ cm; $\tau=100$ fs pulses can be created with a modest magnetic field amplitude of $B_0=3$ mT.

The idea of using resonant deflection cavities in electron microscopy was already researched in the 1970s resulting in picosecond\cite{oldfield1976rotationally} and even sub-picosecond\cite{hosokawa1978generation} electron pulses at GHz repetition rates. After the development of advanced laser-microwave synchronization schemes,\cite{Kiewiet2002} resonant cavities became relevant for laser-triggered pump-probe experiments as well. Especially since the experimental demonstration of a microwave cavity chopping the electron beam of a commercial 200 kV TEM into 3 GHz electron pulses while fully conserving the peak brightness of the original beam.\cite{Verhoeven2018high} Synchronized to a mode-locked laser oscillator, this allows ultrafast time-resolved electron microscopy at the atomic resolution of the original microscope, but at much higher average current. Some additional practical advantages of using microwave deflection cavities for ultrafast electron microscopy are:
\begin{itemize}
    \item No need for an amplified fs laser system to create electron pulses.
    \item Instant switching between continuous mode and pulsed mode.
    \item No fluctuations in electron beam current due to laser pointing instabilities.
    \item The delay of the electron pulse with respect to the laser pulse can be varied by changing the phase of the microwave signal, hence without affecting laser alignment.
    \item Microwave cavities are cheap, robust, reliable, compact and energy-efficient.\cite{Lassise2012compact}
\end{itemize}

Although laser oscillators with repetition rates upto 10 GHz are commercially available,\cite{laserquantum} they are not common. In addition, many laser-triggered processes require relaxation times longer than the typical 300 ps pulse separation in cavity chopped pulsed beams. Ideally one would like to chop the beam into 70 to 90 MHz pulses, the typical repetition rate of most commercial Ti:Sapphire oscillators, enabling pump-probe experiments using a regular Ti:Sapphire oscillator only. This would allow the user to study processes with relaxation times up to 11-14 ns. Dynamic processes requiring longer relaxation times could be studied using a combination of such a microwave cavity and an electrostatic blanker with a rise and fall time of few ns. However, because the resonance frequency of a cavity is inversely proportional to its radius, a 75 MHz cavity would be far too large to fit in the confined space of a electron microscope column. Therefore Lassise \emph{et al.} proposed designing a cavity that supports two orthogonal modes with different resonant frequencies.\cite{Lassise2012miniaturized} By breaking the cylindrical symmetry using an elliptical (or rectangular) cavity, the resonance frequencies of these orthogonal modes can be separated and driven independently. The on-axis magnetic field of such a cavity aligned along the $z$-axis is given by
\begin{equation}
\mathbf{B}=B_1 \sin{(\omega_1 t+\phi_1)} \hat{\mathbf{x}} + B_2 \sin{(\omega_2 t+\phi_2)} \hat{\mathbf{y}}.
\end{equation}
When the $40^\textrm{th}$ and the $41^\textrm{st}$ harmonics of the same 75 MHz laser oscillator are used to create the different microwave signals, yielding resonance frequencies of $\omega_1=2\pi \times 3$ GHz and $\omega_2=2\pi \times 3.075$ GHz respectively, the EM fields in the cavity can be phase-locked to each other \emph{and} to the laser light. As a result, electrons propagating through the dual mode cavity are periodically deflected transversely in the Lissajous pattern of figure \ref{fig:Lissajous}b. By placing a small aperture in a line of this Lissajous beam, it is chopped into ultrashort electron pulses at a repetition rate of 75 MHz allowing pump-probe experiments using a Ti:Sapphire oscillator.

Another important application of sideways deflecting cavities is to spatially resolve temporal information on a detector. Such so-called streak cameras are used for pulse length measurements,\cite{VanOudheusden2010} time-resolved diffraction studies\cite{Musumeci2010} and even time-of-flight Electron Energy Loss Spectroscopy (ToF-EELS).\cite{Verhoeven2016} Note that for these applications the cavity should be placed behind the specimen. At this point it is worth emphasizing the enormous time span that is spatially resolved by the Lissajous pattern. The equation for the temporal resolution of such a streak camera is identical to equation \eqref{eq:tau}, but with the aperture diameter $s$ replaced by the pixel size and $\frac{w}{s}\ll 1$ because the beam would be focused on the screen instead of in the center of the cavity. If a dual mode cavity is placed at a distance $l=10$ cm from a 3 cm $\times$ 3 cm camera with $2000\times2000$  pixels, a magnetic field of $B_0=3$ mT is enough to project a 100 fs time interval on each pixel. The dual mode cavity would then be able to temporally resolve dynamic processes up to 13 ns with 100 fs resolution, corresponding to a huge dynamic range $>10^4$.

\section{Methods}\label{sec:methods}

\begin{figure}
    \centering
    \includegraphics[width=0.5\textwidth]{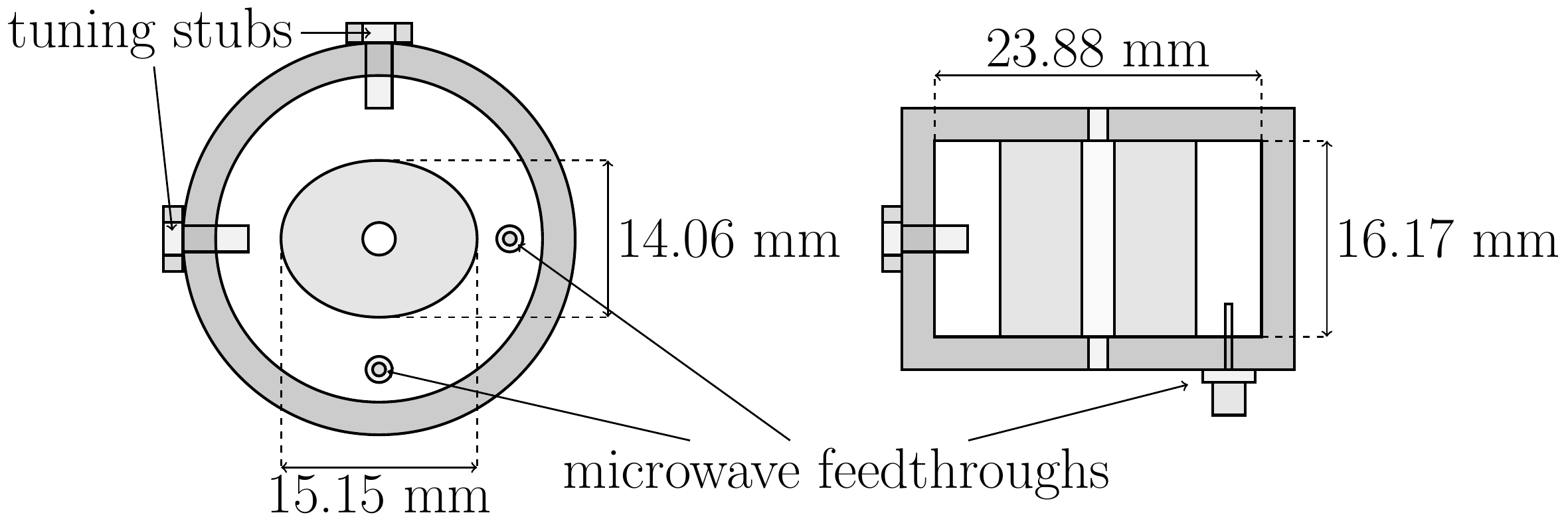}
    \caption{Schematic top view (left) and side view (right) of the dual mode cavity. The eccentricity of the elliptical dielectric material has been exaggerated for clarity.}
    \label{fig:dualmodecavity}
\end{figure}

Figure \ref{fig:dualmodecavity} shows a schematic top view and side view of the dual mode cavity. It consists of a circular copper housing filled with a slightly elliptical piece of dielectric material ZrTiO$_4$, lifting the x,y-degeneracy. The high permittivity and low loss tangent of this material significantly reduces the diameter and power consumption of the cavity.\cite{Lassise2012compact} Furthermore, the cavity contains two antennas to couple in the microwave signals and two metallic tuning stubs to fine-tune the resonance frequencies in both directions.
\begin{figure*}
    \centering
    \includegraphics[width=\textwidth]{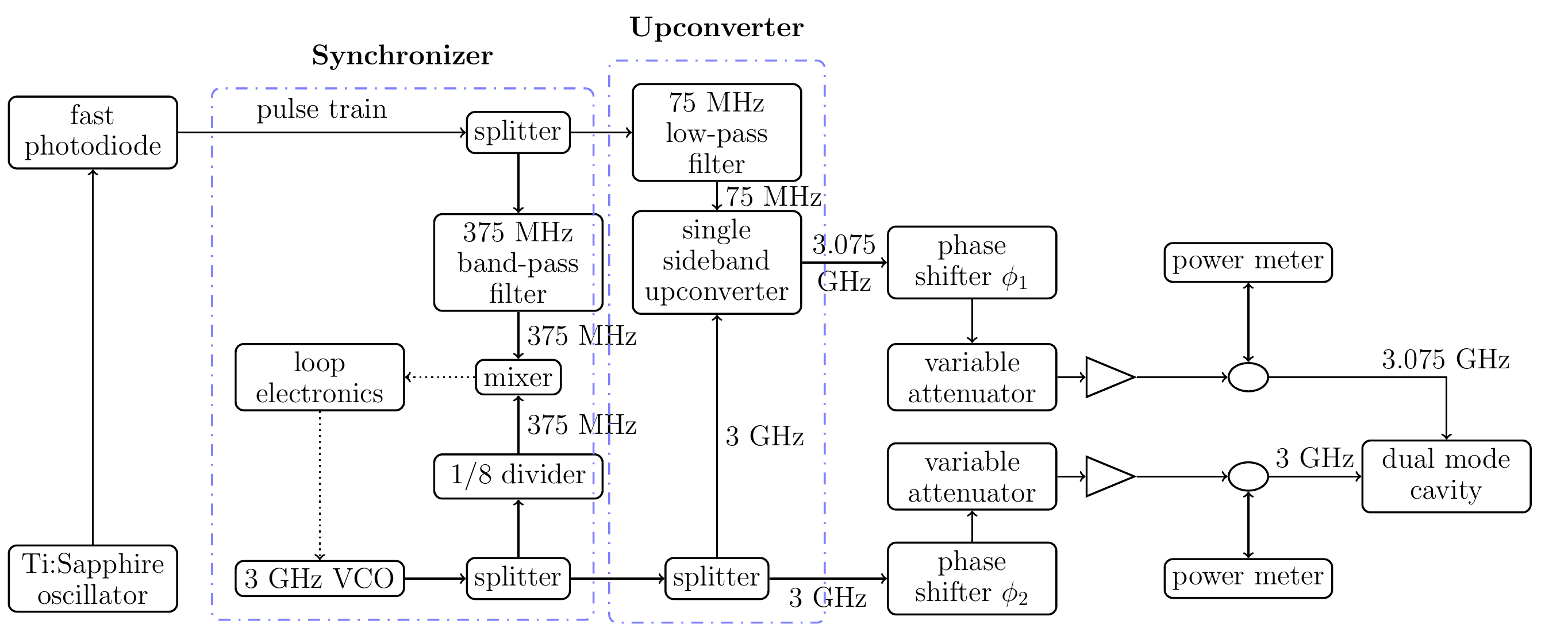}
    \caption{Schematic microwave setup for the dual mode cavity. The 3 GHz signal from a Voltage Controlled Oscillator (VCO) is synchronized to the broadband signal of a Ti:Sapphire laser. Subsequently the synchronized 3 GHz signal is upconverted to 3.075 GHz by mixing it with the 75 MHz component of the laser.}
    \label{fig:RF setup}
\end{figure*}

Figure \ref{fig:RF setup} schematically shows the microwave setup used to generate the phase-locked microwave signals. The left-hand blue dashed box indicates the synchronization system, synchronizing the 3 GHz signal from a Voltage Controlled Oscillator (VCO) to the mode-locked laser pulse train by locking its fifth harmonic to one-eighth of the VCO signal.\cite{Kiewiet2002} This reduces the timing jitter between the laser pulses and the microwave fields in a single mode cavity to below 100 fs.\cite{Brussaard2013} To create the $41^\textrm{st}$ harmonic, the resulting synchronized 3 GHz signal is mixed with the 75 MHz frequency component of the laser pulse train by a single sideband upconverter. This results in a 3.075 GHz signal, phase-locked to the 3 GHz signal and synchronized to the laser pulses. Variable attenuators and the combination of two $2\pi$ 3 GHz phase shifters offer independent control of both the amplitude and phase of each mode in the cavity. Note that laser-triggered electron sources would need a 2-meters-long retroreflecting optical delay line of to cover the 13 ns period, leading to inevitable laser pointing errors.

The cavity is mounted inside a water-cooled copper structure, of which the temperature is measured with an NTC sensor and stabilized to 1 mK using a homebuilt control system. This is necessary to keep the 3 GHz cavity resonance frequency within the 1 MHz control bandwidth of the synchronizer. This structure is inserted in between the C2 aperture (10 $\mu$m) and the minicondenser lens of a modified 200 kV FEI Tecnai TF20, together with a pair of additional deflection coils and a chopping aperture.\cite{Verhoeven2018high} The C2 lens is used to create a crossover in the center of the cavity to eliminate emittance growth, while the minicondenser lens is used to set the illuminated field of view. Standard TEM deflection coils behind the chopping aperture are used to counteract the acquired transverse velocity component due to deflection by the cavity. The SA-aperture is used to block out the second beam, see also the caption of figure \ref{fig:Lissajous}.
\section{Results}\label{sec:results}

\begin{figure}[h]
    \centering
    \includegraphics[width=0.45\textwidth]{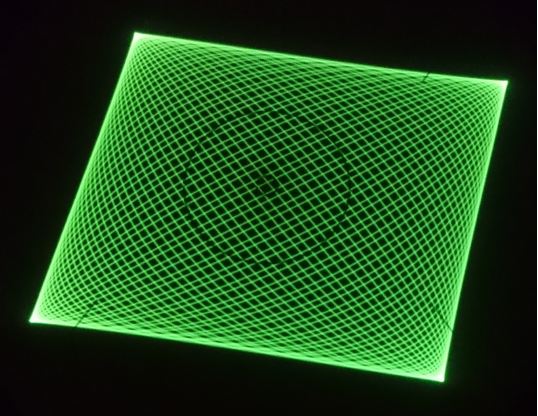}
    \caption{Experimental demonstration of the Lissajous pattern on the fluorescent screen inside the TEM.}
    \label{fig:Lissajousexp}
\end{figure}

To visualize the entire Lissajous pattern, we start at low magnification and low microwave input powers in both cavity modes. The beam is focused onto the fluorescent screen by the C2-lens, resulting in the Lissajous pattern of figure \ref{fig:Lissajousexp}. The fact that the individual lines are distinguishable shows that both modes of the dual mode cavity are phase-locked properly. Changing the phase difference between the two modes results in a change of the Lissajous figure.

\begin{figure*}
 \centering
    \includegraphics[width=\textwidth]{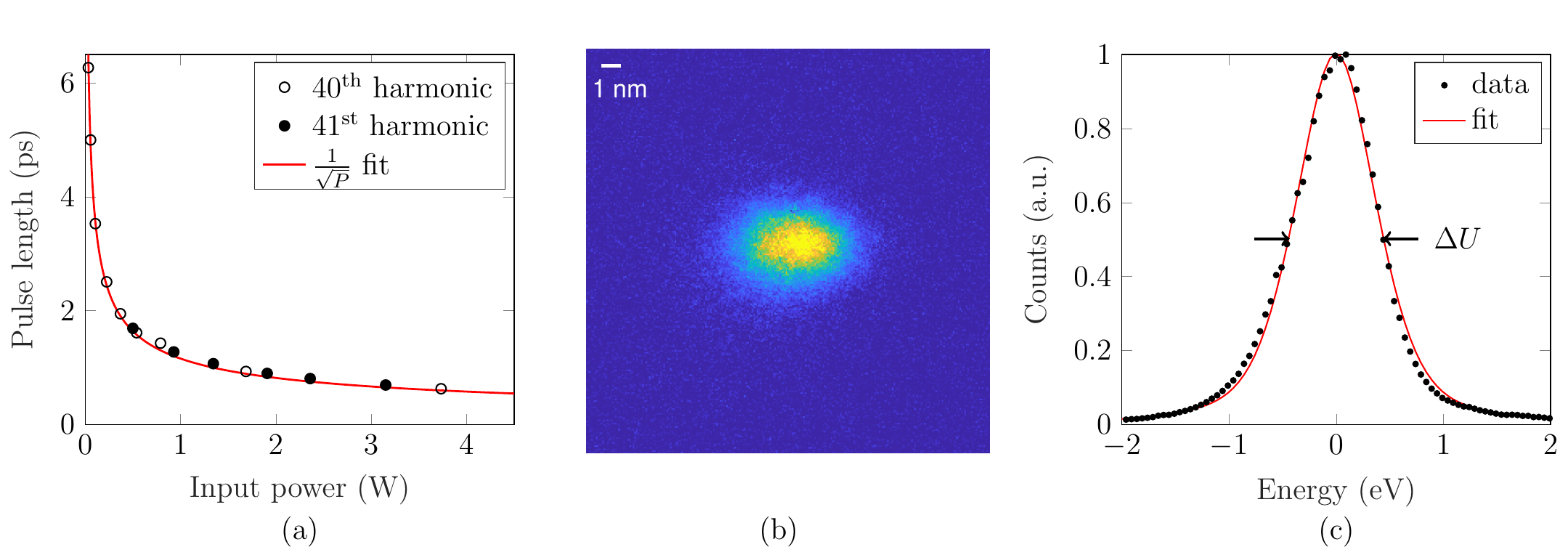}
    \caption{(a) Pulse length as a function of microwave power in the cavity (single mode). The open circles represent the 3 GHz signal, the solid circles represent the 3.075 GHz signal. The red curve is a $1/\sqrt{P}$ fit, calculated for both signals. (b) Image of the 75 MHz pulsed beam focused on the detector for a peak current of $I_p=(0.4\pm0.1)$ nA. The rms size of the focused spot is $\sigma_x=(1.3\pm0.1)$ nm times $\sigma_y=(0.8\pm0.1)$ nm. (c) Measurement of the energy spread of the beam. The solid red curve is a pseudo-Voigt fit with a full width at half maximum of $\Delta U = (0.90\pm0.05)$ eV.}%
    \label{fig:beamquality}%
    \end{figure*}

Then a 10 $\mu$m chopping aperture is inserted, creating a pulsed electron beam. The C2-lens is used to create a crossover in the center of the cavity and the minicondenser lens is used to focus the beam onto the fluorescent screen again. By measuring the decrease in total counts per second on the camera, the pulse length can be determined for varying microwave input power. First, both modes are characterized independently by driving them one at a time. This results in figure \ref{fig:beamquality}a. Both modes show the expected $\tau\propto 1/B\propto 1/\sqrt{P}$ behaviour.\cite{Verhoeven2018high} Note a pulse length of $\tau=(630\pm10)$ fs only requires $(3.7\pm0.1)$ W of microwave input power. The decrease in pulse length with respect to previous measurements \cite{Verhoeven2018high} is explained by using smaller apertures.\cite{VanRens2018}

Then, both orthogonal modes are driven simultaneously to reduce the repetition rate to 75 MHz. Figure \ref{fig:beamquality}b shows an image of the pulsed beam focused onto the detector in which the $40^{\textrm{th}}$ and $41^{\textrm{st}}$ harmonics are driven at $(1.3\pm0.1)$ W and $(1.1\pm0.1)$ W respectively. According to the fit in figure \ref{fig:beamquality}a, the total input power of $(2.4\pm0.1)$ W results in a pulse length of $\tau=(750\pm10)$ fs. The rms size of the focused spot is $\sigma_x=(1.3\pm0.1)$ nm times $\sigma_y=(0.8\pm0.1)$ nm. The slight asymmetry is present in continuous mode as well and is caused by aberrations.  Subsequently the semi-divergence angle is measured in diffraction mode for the same illumination settings, by comparing the diameter of the Ronchigram to the diameter of the first Bragg ring of the diffraction pattern of a gold specimen. This results in a semi-divergence angle of $\alpha = (3.3\pm0.1)$ mrad, corresponding to an rms value of $\sigma_{x'}=\sigma_{y'}=(1.7\pm 0.1)$ mrad for a uniform distribution. For a focused beam the rms normalized transverse emittance in the $j=x,y$-direction can be determined via
\begin{equation}
\epsilon_{n,j}=\beta \gamma \sigma_j \sigma_{j'}, \textrm{~with~} j=x,y,
\end{equation}
in which $\beta=v/c$ is the velocity of the electrons $v$ normalized to the speed of light $c$. Substituting the measured values for $\sigma_x$, $\sigma_y$ and $\sigma_{x'}=\sigma_{y'}$ yields the rms normalized emittance in both directions: $\epsilon_{n,x}=(2.1\pm0.2)$ pm rad and $\epsilon_{n,y}=(1.3\pm0.2)$ pm rad. These measurements were performed at a peak current of $I_p=(0.4\pm0.1)$ nA, i.e. the time-averaged current for the unchopped beam. These measurements allow us to calculate the rms normalized (or reduced) peak brightness
\begin{equation}\label{eq:brightness}
B_{np,\textrm{rms}}=\frac{q}{m c^2}\frac{I_p}{4\pi^2\epsilon_{n,x}\epsilon_{n,y}}=\frac{1}{V^*}\frac{I_p}{2\pi^2 \alpha^2\sigma_x\sigma_y},
\end{equation}
in which $V^*=(1/2+\gamma/2)V$ is the acceleration voltage $V$ multiplied by a relativistic correction term.\cite{Verhoeven2018high} The practical reduced peak brightness as introduced by Bronsgeest \emph{et al.}\cite{bronsgeest2008probe} is then given by $B_{\textrm{pract}}=B_{np,\textrm{rms}}/\ln{2}$. \cite{VanRens2018} Substituting the measured values in equation \eqref{eq:brightness} results in an rms normalized peak brightness of $B_{np,\textrm{rms}}=(7\pm1)\times10^6$ A/m$^2$ sr V. Note that this value presents a lower limit, because the spot size is limited by abberations. A comparison to previously measured values for a continuous beam ($B_{np,\textrm{rms}}=7.5 \times10^6$ A/m$^2$ sr V) and a 3 GHz chopped beam ($B_{np,\textrm{rms}}=6.6 \times10^6$ A/m$^2$ sr V)\cite{Verhoeven2018high} results in the conclusion that the dual mode cavity fully conserves peak brightness. In addition figure \ref{fig:beamquality}c shows the energy spectrum of the 75 MHz electron pulses. The full width at half maximum energy spread obtained from a pseudo-Voigt fit is $\Delta U=(0.90\pm0.05)$ eV, which is equal to the energy spread of the continuous beam as well.\cite{Verhoeven2018high}

\section{Conclusion}\label{sec:conclusion}
We have demonstrated the operation of a dual mode deflection cavity inside a commercial transmission electron microscope to create a beam of ultrashort electron pulses with conservation of peak brightness and energy spread. Sub-picosecond electron pulses are created at a repetition rate of 75 MHz and are synchronized to a Ti:Sapphire oscillator. Furthermore, we have measured rms normalized transverse emittances of $\epsilon_{n,x}=(2.1\pm0.2)$ pm rad and $\epsilon_{n,y}=(1.3\pm0.2)$ pm rad, for a peak current of $I_p=(0.4\pm0.1)$ nA. This results in a rms normalized peak brightness of $B_{\textrm{rms}}=(7\pm1)\times10^6$ A/m$^2$ sr V, demonstrating that deflecting electrons using multiple modes in a microwave cavity does not affect the peak brightness of the beam. Also the energy spread $\Delta U = (0.90\pm 0.05)$ eV is unaffected by the dual mode cavity. This allows for atomic spatial resolution and sub-ps temporal resolution pump-probe experiments in which a standard Ti:Sapphire oscillator can be used to excite dynamic processes.

In addition, the experimental demonstration of the Lissajous beam, shows that the dual mode cavity can be used as a streak camera that can resolve temporal differences up to 13 ns with sub-ps resolution, corresponding to a dynamic range $>10^4$. As a final remark, this demonstration of controlling the motion of electrons by driving multiple modes in a single cavity, opens the door to even more advanced pulsed beam manipulation using microwave cavities. For example, a microwave cavity that supports multiple higher harmonics of the same mode, could extend the linear regime of the cavity electromagnetic fields, allowing better temporal compression or higher energy resolution in TM$_{010}$ mode cavity based time-of-flight EELS setups.\cite{Verhoeven2017time}


\begin{acknowledgments}
The authors would like to thank Eddy Rietman, Ad Kemper, Harry van Doorn and Iman Koole for their invaluable technical support. This work is part of an Industrial Partnership Programme of the Netherlands Organisation for Scientific Research (NWO).
\end{acknowledgments}

\bibliography{mybibliography}
\end{document}